\documentclass{ifacconf}
\usepackage{listings}
\renewcommand*\thelstnumber{\oldstylenums{\the\value{lstnumber}}}
\usepackage{amsmath}
\usepackage{amssymb}
\usepackage{arydshln}
\usepackage{csvsimple}
\usepackage{colortbl}
\usepackage{multicol}
\usepackage{algorithm}
\usepackage[noend]{algpseudocode}
\usepackage{microtype}
\usepackage{times}
\usepackage{prettyref} %
\newrefformat{fig}{Fig.~\ref{#1}}
\newrefformat{tab}{Table~\ref{#1}}

\usepackage{layouts}

\usepackage{graphicx}      %
\usepackage{natbib}        %
\usepackage[dvipsnames]{xcolor}
\usepackage{todonotes}
\newtheorem{definition}{Definition}[section]

\definecolor{DarkGreen}{rgb}{0.0, 0.5, 0.0}
\definecolor{BurntOrange}{rgb}{0.8, 0.33, 0.0}
\definecolor{LightCyan}{rgb}{0.88,1,1}

\definecolor{OliveGreen}{rgb}{0.33,0.42,0.18}
\lstdefinestyle{MEL}{
    tabsize=2,
    captionpos=b,
    xleftmargin=0pt,
    framexleftmargin=5pt,
    keywordstyle=\color{blue}\bf,
    comment=[l]{//},
    morecomment=[s]{/*}{*/},
    commentstyle=\color{OliveGreen}\it,
    stringstyle=\color{red},
    breaklines=true,
    showstringspaces=false,
    basicstyle=\footnotesize,
    emph=[1]{module,real,boolean,initial,constant},
    emphstyle=[1]{\color{magenta}},
    emph=[2]{if,when,then,else,end,invariant,trans,sum,foreach,in,do,done},
    emphstyle=[2]{\color{blue}},
    emph=[3]{der},
    emphstyle=[3]{\textbf}}%
    
\newcommand{\mmDM}{\texttt{mmDM}}

\def\BibTeX{{\rm B\kern-.05em{\sc i\kern-.025em b}\kern-.08em
		T\kern-.1667em\lower.7ex\hbox{E}\kern-.125emX}}

\newcommand{\ffalse}{\mathbf{F}}
\newcommand{\ttrue}{\mathbf{T}}

\usepackage{xspace} %

\def\Ms{\text{$M_s$}\xspace} %
\def\MB{\text{$M_b$}\xspace} %

\newcommand\vsm[1]{\text{$\textit{v}_{\text{sm},#1}$}\xspace} %

\def\vout{\text{$\textit{v}_{\text{pack}}$}\xspace} %
\def\iout{\text{$\textit{i}_{\text{pack}}$}\xspace} %

	\def\Sa{\text{$\text{S}_{1}$}\xspace} %
	\def\Sb{\text{$\text{S}_{2}$}\xspace} %
	\def\Sc{\text{$\text{S}_{3}$}\xspace} %
	\def\Sd{\text{$\text{S}_{4}$}\xspace} %
	
	\newcommand{\forw}{\mathsf{forward}}
	\newcommand{\back}{\mathsf{backward}}

	\newcommand{\by}{\mathsf{bypass}}

	\newcommand{\vcell}[1]{\text{$\textit{v}_{\text{cell,\(#1\)}}$}\xspace} %
	\newcommand{\icell}[1]{\text{$\textit{i}_{\text{cell,\(#1\)}}$}\xspace} %

	\newcommand{\fltcell}[1]{\text{\(f_{\text{cell},#1}\)}\xspace} %

\renewcommand{\cite}{\citep}

\begin{document}
\begin{frontmatter}

\title{Fault Diagnosability Analysis of Multi-Mode Systems}

\author{Fatemeh Hashemniya*}, 
\author{Beno{\^i}t Caillaud**}, 
\author{Erik Frisk*},
\author{Mattias Krysander*},  \textbf{and}
\author{Mathias Malandain**}

\address[First]{Department of Electrical Engineering\\ 
	Linköping University, SE 581-83, Linköping, Sweden\\ 
	e-mail: \{fatemeh.hashemniya, erik.frisk, mattias.krysander\}@ liu.se}
	
\address[Second]{National Institute for Research in Digital Science and Technology (Inria),
		Inria centre at Rennes University, Rennes, France.\\ 
		e-mail: \{benoit.caillaud, mathias.malandain\}@inria.fr}

\begin{abstract}                %
Multi-mode systems can operate in different modes, leading to large numbers of different dynamics. Consequently, applying traditional structural diagnostics to such systems is often untractable. To address this challenge, we present a multi-mode diagnostics algorithm that relies on a multi-mode extension of the Dulmage-Mendelsohn decomposition. We introduce two methodologies for modeling faults, either as signals or as Boolean variables, and apply them to a modular switched battery system in order to demonstrate their effectiveness and discuss their respective advantages.
\end{abstract}

\end{frontmatter}

\section{Introduction}

Fault detection and diagnosis are important for the health monitoring of physical systems.
Model-based approaches for single-mode, smooth, systems is a well-established field, supported by a  large body of literature covering various approaches like structural methods~\cite{Blanke2006}, parity space techniques, and observer-based methods~\cite{isermann2006fault}.

While single-mode systems are often described using differential algebraic equations (DAEs), the modeling of non-smooth physical systems yields switched DAEs, also known as multi-mode DAEs (mmDAEs), which combine continuous behaviors, defined as solutions of a set of DAE systems,
with discrete mode changes~\cite{Trenn2012,benveniste:hal-03045498}. Direct application of
traditional fault diagnosis methods to all possible configurations of multi-mode systems quickly
becomes intractable, as the number of modes tends to be exponential in the size of the system.
The method proposed by \cite{khorasgani2017structural} works around this issue by coupling a mode estimation algorithm with a single-mode diagnosis methodology, akin to just-in-time compilation in computer science. This approach unfortunately puts the burden on solving mode estimation problems, which often turn out to be intractable for the same reason. 

Structural fault detectability and isolability is a graph-based method to evaluate diagnosability properties on
DAEs~\cite{frisk2012diagnosability}. It is based on the Dulmage-Mendelsohn decomposition (DM), a building block of the
structural analysis of equation systems. In this study, we show how its extension to multi-mode
systems, introduced in~\cite{benveniste:hal-03768331}, can be applied in the context of structural
fault detectability and isolability of mmDAEs~\cite{Hashemniya2023MM}. Building upon our previous research
studies, the methods presented in this paper represent advancements in diagnostic methodologies for
multi-mode systems, providing novel ways to study the diagnosability of multi-mode systems
without enumerating their modes.

The case study used throughout this article is a model of a reconfigurable battery system, in which
switching strategies enable to produce an AC output without relying on a central
inverter~\cite{balachandran2021design}. This model is parametrized by the number of battery cells,
so that both the inherent complexity associated with the diagnostics of such systems and the
scalability of our approaches can be addressed.

\section{Problem formulation}

The main problem studied here is how to perform the diagnosability analysis of multi-mode systems.
Specifically, the method should be able to determine whether a fault $f_i$ can be detected and/or isolated from another fault $f_j$, and if so, under which operational modes.

As an illustrative example, we consider throughout this paper a battery pack composed of $n$ switched submodules (SM) arranged in series, as introduced in \cite{paper_with_Arvind}. Each SM has four valid modes, resulting in a total of $4^n$ system configurations that need to be investigated to calculate the multi-mode isolability performance matrix introduced in \cite{Hashemniya2023MM}. 
There, individual configurations have different diagnosability properties leading to complexity concerns, since the number of configurations grows exponentially with the number of SMs.
This inherent computational complexity is a key problem and motivating factor of this study.

As discussed in \cite{paper_with_Arvind}, even a modest-sized example with a limited number of SMs results in a large number of system configurations, making any brute-force approach untractable when faced with realistic problem sizes. A more efficient method for assessing structural diagnosability is needed. In \cite{paper_with_Arvind}, an approach was introduced to reduce the number of modes and system configurations, resulting in complexity gains. However, that solution was designed specifically for that particular use case. To establish a comprehensive diagnostic framework applicable to general multi-mode systems, a generalized multi-mode Dulmage-Mendelsohn (mmDM) decomposition \cite{benveniste:hal-03768331} is used to develop a method for general multi-mode diagnosability analysis. In this framework, the modeling of the faults themselves is an important question: two different approaches are presented here, and their properties are studied.

\section{Single-mode diagnosis}
\label{sec:diagnosis}

This section briefly recaps the foundations of a graph theoretical method and corresponding definitions for single-mode detectability and isolability analysis. These concepts will be extended in later sections to the multi-mode case. 

\begin{figure}[htbp]
	\centering
	\includegraphics[width=.40\columnwidth]{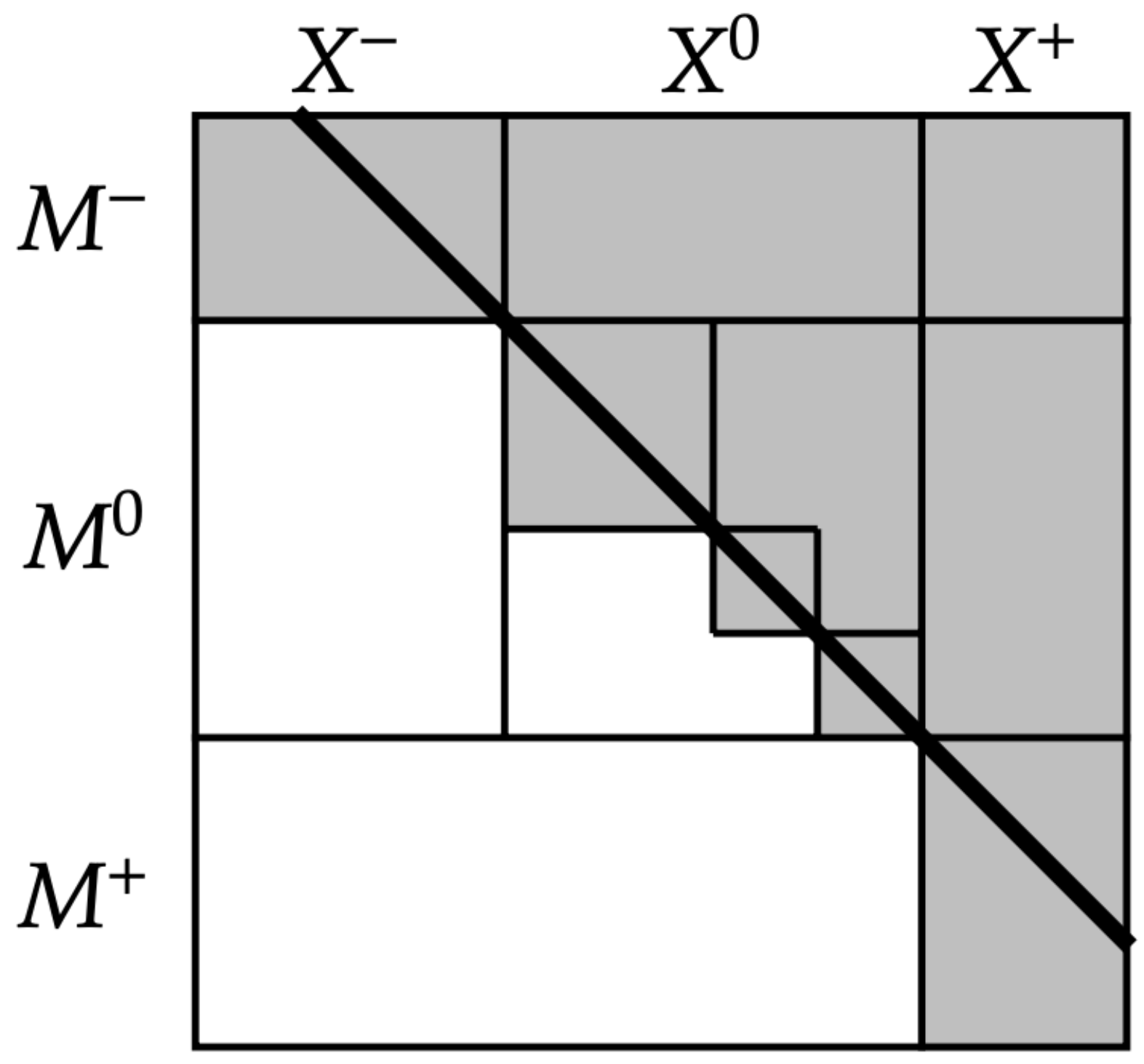}
	\vspace{-2mm}
	\caption{Dulmage-Mendelsohn decomposition.}\label{fig:dmperm}
\end{figure}

The Dulmage-Mendelsohn decomposition~\cite{Dulmage1958} is a basic theoretical tool, illustrated by Figure~\ref{fig:dmperm}, where the structure of a set of equations $M$ with the set of unknown variables $X$ is represented as an incidence matrix. Three main parts of $M$ can be identified in the decomposition: $M^-$ is called the structurally underdetermined part, $M^0$ is the structurally well-determined part, and $M^+$ is the
structurally overdetermined part. %

The overdetermined part has more equations than unknowns, which implies that there exists some degree of redundancy.
Assume that each fault $f$ is present in one equation, denoted as $e_f$, then the following definitions of structural detectability and isolability from \cite{frisk2012diagnosability} will be used.
\begin{definition}\label{def:str_det}
	A fault $f$ is structurally detectable in a model $M$ if $e_f\in
	M^+$.
	A fault $f_i$ is structurally isolable from $f_j$ in a model $M$ if
	  $e_{f_i} \in (M\setminus \{e_{f_j}\})^+$. \hfill $\blacklozenge$
\end{definition}
Fault detectability and isolability of a model will be called the diagnosability of the model. From Definition~\ref{def:str_det}, it appears that one DM decomposition is required to compute the detectability of a given fault or the isolability of one fault from another fault. For a multi-mode system, detectability and isolability can be computed in the same way for each mode separately, but as said before,
this naive way of computing diagnosability properties quickly becomes intractable both in terms of computational complexity and memory usage. In this work, we propose to use a multi-mode framework where modes are encoded efficiently to avoid the need for a complete enumeration of system modes and to use a multi-mode version of the DM decomposition that can be used to compute detectability or isolability properties for all modes in one execution.

\section{Multi-mode Systems} \label{sec:mmsystems}

This section describes the modeling framework for multi-mode systems, including two distinct methodologies for modeling faults. This multi-mode modeling framework will be exemplified in a case study of a reconfigurable battery system with integrated converters.
While this example may seem simple, it is complex enough to illustrate the difficulties inherent to the fault diagnosis of multi-mode systems.
Note that the method presented in this paper can easily be extended to handle more complex, non-linear, multiphysics models.

\begin{figure}[t!]
	\centering
	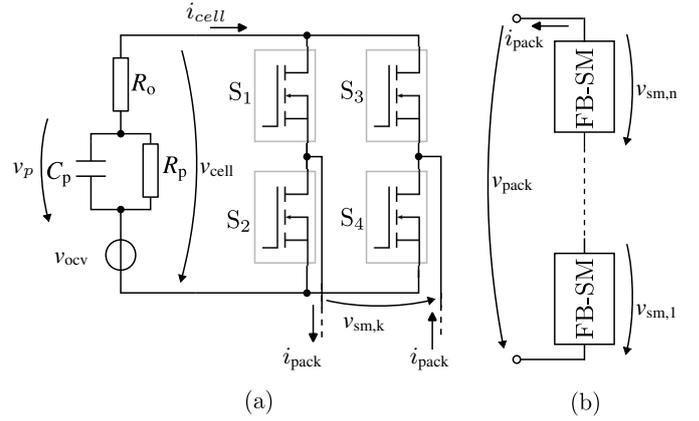
	\vspace{-7mm}
	\caption{Reconfigurable battery system with integrated converters. (a) Battery pack with $n$ submodules. (b) Battery submodule.}
	\label{fig:BIMMC_FBSM}
\end{figure}

The modular battery pack considered includes $n$ SMs in series, numbered from 1 to $n$, as illustrated in Fig.~\ref{fig:BIMMC_FBSM}. SM $k$ in Fig.~\ref{fig:BIMMC_FBSM}(a) includes a battery cell, modeled as an equivalent circuit model, along with a full-bridge converter consisting in four MOSFET switches, i.e., S$_1$-S$_4$, and two sensors, i.e., a voltage and a current sensor, measuring $ \vcell{k}$ and $\icell{k}$ respectively. Additionally, two global pack sensors measure the output voltage \vout and the output current \iout respectively, as shown in Fig.~\ref{fig:BIMMC_FBSM}(b). Faults for each battery cell and each sensor will be considered, and we denote by $\mathcal{F}$ the set of faults.

The mode of an SM is given by the states of its four switches, i.e., there are in total $2^4$ switch configurations. However, many of these switch configurations lead to short circuit; in practice, there are only 4 modes that are used. Table~\ref{tab:SMswmodes} shows the four valid operation modes for each SM based on the states of the switches. It would be possible to model the modes of an SM using the states of the 4 switches, but here, a more compact description will be used that also reduces the number of Boolean variables. In this approach, the Boolean variables $\forw$ and $\back$ encode the four valid modes as shown in Table~\ref{tab:SMswmodes}. With this abstraction, the two bypass modes are modeled identically, and there is indeed no need to distinguish them. Furthermore, remark that the mode $\forw = \ttrue$, $\back = \ttrue$ is forbidden. %
\begin{table}[h!]
	\centering
	\caption{Full-bridge sub-module modes.}
	\vspace*{-3mm}
	\label{tab:SMswmodes}
	\begin{tabular}{c@{\hspace{3pt}}c|l|@{\hspace{3pt}}c@{\hspace{3pt}}|@{\hspace{3pt}}c@{\hspace{3pt}}|@{\hspace{3pt}}c@{\hspace{3pt}}|@{\hspace{3pt}}c@{\hspace{3pt}}|@{\hspace{3pt}}c}
		\multicolumn{2}{c}{encoding} & mode &\multicolumn{4}{c}{switches} & \, \\
		$ \forw $ & $ \back $ & & \Sa & \Sb & \Sc & \Sd & \vsm{k}\\  
		\hline
		$\ttrue $ & $\ffalse $ & forward   &on  & off & off & on  & \vcell{k}\\
		$\ffalse $ & $\ttrue $ & backward  &off & on  & on  & off  & $-$\vcell{k}\\
		$\ffalse $ & $\ffalse $ & bypass 1  &on  & off & on  & off & 0 \\
		$\ffalse $ & $\ffalse $ & bypass 2 &off & on  & off & on  & 0\\
	\end{tabular}
\end{table}

The code for defining the multi-mode Differential Algebraic Equation (mmDAE) of an SM is:
\begin{lstlisting}[style=MEL, label=cell]
	module SM()
	// Mode variables
	forward: boolean; 
	backward: boolean; 
	invariant !(forward & backward);
	// Faults
	constant f_cell : real;
	constant f_i_cell : real;
	constant f_v_cell : real;
	// Equations of a cell
	e1 : v_p_der = i_cell / Cp - v_p / (Rp * Cp);
	e2 : v_cell = v_p + R0 * i_cell + v_ocv + f_cell;
	e3 : v_p_der = der(v_p);
	e4 : v_sm =
	     if forward then v_cell else
	     if backward then - v_cell
	     else 0.;
	e5 : i_cell =
	     if forward then i_pack else
	     if backward then - i_pack
       	 else 0.;
	e6 : y_i_cell = i_cell + f_i_cell;
	e7 : y_v_cell = v_cell + f_v_cell;
	end
\end{lstlisting} 
The model is expressed in the MEL equation language presented in~\cite{benveniste:hal-03768331}. The model begins with a declaration of all variables, then the equations are defined. Three types of variables need to be distinguished in a model used for diagnosability analysis. Mode variables are declared as Booleans, unknowns as real variables, and everything else, including model parameters, sensor signals, and fault signals, as real constants. Here, only the declaration of the mode variables and fault signals are shown. Note that the invariant declaration (line $5$) states that the mode assignment $\forw = \ttrue$, $\back = \ttrue$ is not valid. 

Equations $ \text{e1-e3} $ outline the behavior of the cell, $ \text{e4-e5} $ govern the switches, and $ \text{e6-e7} $ address the cell current and cell voltage sensors. This model is indeed a multi-mode model, as equations  $ \text{e4-e5} $ depend on the operating mode of the system. The fault modeling strategy itself will be discussed separately at the end of this section.

To expand the system to $n$ SMs within the battery pack, and subsequently integrate it by incorporating pack sensors, the following statements are included in model $\Ms$:
\begin{lstlisting}[style=MEL]
// System architecture
#include "SM.mel"
g1 : v_pack = sum { k in 1 .. N : c[k].v_sm };
foreach k in 1 .. N do
  g2[k] : i_pack = c[k].i_pack;
done;
g3 : y_i_pack = i_pack + f_i_pack;
g4 : y_v_pack = v_pack + f_v_pack
\end{lstlisting}
where equation $ \text{g1} $ describes the output voltage of the pack, $ \text{g2[k]} $ states current relationships, and $ \text{g3-g4} $ correspond to the pack's current and voltage sensors, respectively.

\subsection{Fault signal approach}
\label{sec:fault-signal-approach}

Faults can be modeled either as real signals or as Boolean mode variables. 
In the \emph{fault signal} approach, a fault is modeled as a real variable $f$ such that $f\neq 0$ in the presence of the fault, $f = 0$ otherwise. The model with fault signals is denoted $\Ms$. In it, the considered faults of an SM are modeled with the fault signals $f_{\text{cell}}$ for a general cell fault, $f_{i_{\text{cell}}}$ for a current sensor fault, and $f_{v_{\text{cell}}}$ for a voltage sensor fault. Furthermore, sensor faults measuring pack current and voltage are modeled with fault signals $f_{i_{\text{pack}}}$ and $f_{v_{\text{pack}}}$ respectively. Therefore, the total number of faults is $3n+2$, where $n$ is the number of SMs in the pack. 

\subsection{Boolean fault approach}
\label{sec:boolean-fault-approach}

Faults can also be considered to be discrete modes of the system, i.e., fault modes. In this approach, faults are modeled by Boolean mode variables.
As a result, equations containing faults also need to be changed, resulting in the following declarations:
	\begin{lstlisting}[style=MEL] 
		F_cell : boolean;
		F_i_cell : boolean;
		F_v_cell : boolean;
		if !F_cell then e2 : v_cell = v_p + 
	                     R0 * i_cell + v_ocv end;
		if !F_i_cell then e6 : y_i = i_cell end;
		if !F_v_cell then e7 : y_v = v_cell end;
		if !F_i_pack then g3 : y_i_pack = i_pack end;
		if !F_v_pack then g4 : y_v_pack = v_pack end
	\end{lstlisting}
	
For example, the last equation describes that the voltage sensor is measuring the correct voltage if the sensor is not faulty. 
This model will be called $\MB$, where subscript $b$ stresses the fact that faults are modeled with Boolean variables.

	One difference between the two methodologies lies in the number of Boolean variables, which directly impact the computational complexity, as we shall see. The first approach involves a total of $2n$ Boolean mode variables, where $n$ is the number of SMs, whereas the second approach includes $5n + 2$ Boolean variables.
	Each approach possesses its respective advantages and drawbacks, which will be discussed in Section~\ref{Results}.

\section{Multi-mode Dulmage-Mendelsohn Decomposition} \label{sec:mmdm}

\newcommand{\bool}{\mathbb{B}}

\newcommand{\mmatch}{\mathcal{M}}

\newcommand{\squared}{0}
\newcommand{\underdet}{-}
\newcommand{\overdet}{+}
\newcommand{\eover}[1]{\varepsilon^\overdet_{#1}}
\newcommand{\xover}[1]{\chi^\overdet_{#1}}

\newcommand{\fsquared}{\varphi^\squared}
\newcommand{\funderdet}{\varphi^\underdet}
\newcommand{\foverdet}{\varphi^\overdet}

\newcommand{\eof}{\mathcal{E}}
\newcommand{\xof}{\mathcal{X}}
\newcommand{\emof}{\mathcal{E}^{-1}}
\newcommand{\xmof}{\mathcal{X}^{-1}}

\newcommand{\bcond}{\gamma}

Multi-mode structural diagnosis analysis requires an efficient way to compute the DM decomposition of a multi-mode system.
In this section, we present the necessary background about the DM decomposition, then introduce a multi-mode extension that does not require enumerating the modes of the system.

\subsection{The single-mode case}
\label{sub:single-mode-dm}

The DM algorithm, introduced
in~\cite{Dulmage1958}, is a canonical decomposition of the set of
vertices of a bipartite graph $G = (E \uplus X, A \subseteq E \times X)$
that is commonly used for solving systems of algebraic equations.
When applied to any bipartite graph $G$,
this decomposition uniquely partitions set $E$ (respectively, set $X$) into
three subsets $E^\underdet$, $E^\squared$ and $E^\overdet$ (resp.,
$X^\underdet$, $X^\squared$ and $X^\overdet$).
As stated in Section~\ref{sec:diagnosis},
these subsets take a specific meaning when applied to the adjacency graph of an algebraic system of equations $M$.

In what follows, we abstract model $M$ by its \emph{adjacency graph},
where each vertex in $E$ represents an
equation, each vertex in $X$ represents a variable, and an edge
$(e,x)$ is in $A$ if and only if variable $x$ occurs in equation $e$.
The DM decomposition then partitions the set of equations $E$ into three subsystems: we say
that $E^\underdet$ is the \emph{underdetermined} part of $E$,
$E^\squared$ is its \emph{square} (or \emph{well-determined}) part,
and $E^\overdet$ is its \emph{overdetermined} part; the corresponding
subsets of $X$ are the subsets of their dependent variables.
We only focus here on the overdetermined parts $E^\overdet$ and $X^\overdet$, as these are the ones to be used for fault analysis.

An efficient algorithm for computing the DM decomposition of sparse
systems was published by~\cite{Pothen1990}. A
maximum matching $\mmatch$ of the system's adjacency graph is required
as an input; the resulting decomposition is independent from the choice of a particular maximum matching.

Let $E^u$ (respectively, $X^u$) be the set of
equations (resp., variables) that are not matched in $\mmatch$. Then, $E^\overdet$ and $X^\overdet$ are the subsets of $E$ and $X$ (respectively) that are reachable via an alternating path from $E^u$, where an \emph{alternating path} is a path whose edges belong alternatively to $\mmatch$ and
$A \setminus \mmatch$.

\subsection{Multimode extension}
\label{sub:multi-mode-dm}
This section introduces {\mmDM}, a multimode generalization of the DM algorithm,
presented in~\cite[Section~4.2]{benveniste:hal-03768331}. It is
designed for algebraic systems of equations in which both equations
and variables can be enabled or disabled depending on the
satisfaction of a set of Boolean conditions on mode
variables. Equations may also contain \emph{if-then-else} operators,
meaning that the incidence relation may also vary depending on the
satisfaction of Boolean conditions on the mode variables.
In what follows, we denote by $\mathbb{M}$ the set of valid modes, i.e., the set of all valuations of the mode variables satisfying the invariant of the model as introduced in Section~\ref{sec:mmsystems}.

The key principle of this algorithm is that it is based on a \emph{dual
representation} of the structure of the system: each vertex and
edge of the incidence graph is labeled by a Boolean condition that
characterizes the set of modes in which it is
enabled.
Hence, we assume in the sequel that we are given Boolean functions
$\bcond_e, \bcond_x, \bcond_a : \mathbb{M} \rightarrow \bool$ for every
$e\in E$, $x\in X$ and $a\in A$. Given an edge $a = (e,x) \in A$,
denote $\eof(a) = e$ its incident vertex in $E$ and $\xof(a) = x$ its
incident vertex in $X$. The inverse mappings
$\emof(e) = \{ (e,x) \in A \}$ (resp., $\xmof(x) = \{ (e,x)\in A \}$)
define the set of edges that are incident to vertex $e\in E$
(resp., $x\in X$).

The choice of one maximum matching per mode is performed without
enumerating the modes via an adaptation of the maximum weight
perfect matching algorithm detailed in~\cite{benveniste:hal-03768331},
Section~4.3.  This maximum matching is given by its
characteristic functions
$T_a : \mathbb{M} \rightarrow \bool$ for all edges $a\in A$.

For each equation $e \in E$, a function $\eover{e} : \mathbb{M} \rightarrow \bool$ is defined, whose final
value states the modes in which this equation belongs to the
overdetermined part of model $M$.
In a similar fashion, a function $\xover{x} : \mathbb{M} \rightarrow \bool$ is defined for each
variable $x \in X$.

Each $\eover{e}$ is initialized so that it encodes the set of modes
in which equation $e$ is unmatched, that is:
\begin{equation}
  \begin{array}{r@{~\leftarrow~}l}
    \eover{e} & \bcond_e \land \lnot \left( \bigvee_{a \in \emof(e)} \; T_a \right)
  \end{array}
\label{eq:overdet-init}
\end{equation}
while function $\xover{x}$ is initialized to
$\ffalse$, the \emph{false} constant.

The \emph{propagation steps} that follow consist in
exploring alternating paths from the ``overdetermined sets''
$\eover{e}$, $\xover{x}$ and updating the corresponding functions as follows, for all $x \in X$ and all  $e \in E$:
\begin{equation}
\begin{array}{r@{~\leftarrow~}l}
  \xover{x} & \xover{x} \lor \left( \bcond_x \land \bigvee_{a \in \xmof(x)} \left( \lnot T_a \land \bcond_a \land \eover{\eof(a)} \right) \right) \\
  \eover{e} & \eover{e} \lor \left( \bcond_e \land \bigvee_{a \in \emof(e)} \left( T_a \land \xover{\xof(a)} \right) \right)
\end{array}
\label{eq:propagation}
\end{equation}
until a fixpoint is reached.

An OCaml implementation of this method, based on Binary Decision Diagrams (BDD, see \cite{Bryant1986}), is used for the multi-mode fault diagnosability analyses presented next.

\section{Multi-mode diagnosability}

We recall that the structural diagnosability analysis of a single-mode system is performed by computing the detectability of every fault $f\in \mathcal{F}$ and the isolability for every fault pair $(f_i,f_j)\in \mathcal{F}\times \mathcal{F}$. Note that, from Definition~\ref{def:str_det}, detectability of a fault $f$ can be seen as the isolability of $f$ from the empty set of faults, i.e., the no-fault case \texttt{NF}. Thus, the diagnosability
of a single-mode model $M$ can compactly be described by%
\begin{equation}\label{diag}
	\begin{split}
		D_{(f, \texttt{NF})}^{M} &= \ttrue \text{ iff } f \in \mathcal{F} \text{ is detectable} \\ 
		D_{(f_i, f_j)}^{M} &= \ttrue \text{ iff } (f_i,f_j) \in \mathcal{F}\times \mathcal{F} \text{ is isolable.}
	\end{split}
\end{equation}
Extending diagnosability analysis to a multi-mode model, with valid modes $\mathbb{M}_s$, then amounts to computing a Boolean function $D_{(f_i, f_j)} : \mathbb{M}_s \rightarrow \bool$,
for every $f_i\in \mathcal{F}$ and $f_j\in \mathcal{F} \cup \{\texttt{NF}\}$, such that, for any valid assignment of the mode variables $S=S_0$:
\[
	D_{(f_i, f_j)}[S:=S_0] \; = \; D_{(f_i, f_j)}^{M[S:=S_0]}
\]
where $M[S:=S_0]$ is the single-mode model obtained from $M$ by setting its mode variables according to $S_0$.

Next, we present algorithms for computing the diagnosability function, for both fault modeling methodologies, by considering all valid modes at once (that is, without enumerating them).

\subsection{Diagnosability based on fault signals}

In the fault signal approach, the set $\mathcal{F}$ of faults in a multi-mode model $\Ms$ is described by the fault signals $f = (f_1, f_2, \ldots, f_n)$ such that $f_i \neq 0$ if fault $i$ is present, $f_i = 0$ otherwise. The set of Boolean variables $S = (S_1, \ldots, S_n)$, called \emph{system mode variables}, represent, e.g., valve and switch positions in the system. It is assumed that faults are introduced only in equations without system mode variables.%

For each equation $e \in E$ in $\Ms$, {\mmDM} returns a Boolean function $\eover{e} : \mathbb{M}_s \rightarrow \bool$, where the superscript $+$ denotes the overdetermined part, and $\mathbb{M}_s$ denotes the set of valid modes. These functions will be used to answer %
diagnosability questions. The \texttt{dd} library~\cite{tulip-control:dd} is used for efficiently handling these functions as BDDs.%

\begin{definition}[Structural detectability]
	A fault $f_i$ is structurally detectable in a multi-mode system $\Ms$ if 
	\begin{equation}
		\eover{e_{f_i}}(\Ms) \not \equiv \ffalse \ , 
	\end{equation}
	where $e_{f_i}$ is the equation including fault $f_i$. \hfill $\blacklozenge$
\end{definition}
This means that a fault is structurally detectable if there exists at least one valid mode assignment $S=S_0$ such that $\eover{e_{f_i}}(\Ms)[S:=S_0] = \ttrue$. Given a set of modes $\mathbb{S}$, fault $f_i$ is detectable in all modes belonging to $\mathbb{S}$ iff $1_{\mathbb{S}}\rightarrow\eover{e_{f_i}}(\Ms)$ is the constant true function. Hence, a fault $f_i$ is detectable in all valid modes iff
	$( 1_{\mathbb{M}_s} \rightarrow \eover{e_{f}}(\Ms) ) \; \equiv \; \ttrue$.%

Consider a system with only one SM, as defined in Section~\ref{sec:mmsystems}, and fault $f_{\text{cell}}$. This fault is detectable for the modes defined by 
$\eover{e_{f_{\text{cell}}}}(\Ms) = \lnot \back \lor \lnot \forw$, which is the model invariant $1_{\mathbb{M}_s}$. Hence, fault $f_{\text{cell}}$ is detectable in all valid modes. 	

Algorithm~\ref{alg:mmdetectability} computes the detectability of all faults in model $\Ms$. In line~\ref{row:always-det}, the Boolean expression is simplified for faults that are detectable in all valid modes. Note that only one run of $\texttt{mmDM}$ is performed to compute the detectability of all faults in all modes. This algorithm will be used when computing fault isolability.%
\begin{algorithm}
	\caption{$\langle D_{(f,\texttt{NF})} \rangle_{f\in \mathcal{F}} = \text{Detectability}(\Ms)$
	}\label{alg:mmdetectability}
	\begin{algorithmic}[1] %
	\State $\langle \eover{e}(\Ms) \rangle_{e\in E} = \texttt{mmDM}(\Ms)$
	\ForAll{$f \in \mathcal{F}$}		
		\If{$1_{\Ms} \rightarrow \eover{e_f}(\Ms) \; \equiv \; \ttrue$}
			\State $D_{(f,\texttt{NF})} = \ttrue$ \Comment{Detectable in all valid modes} \label{row:always-det}
		\Else 	
			\State $D_{(f,\texttt{NF})} = \eover{e_{f}}(\Ms)$
		\EndIf
	\EndFor		
	\end{algorithmic}
\end{algorithm}

\begin{definition}[Structural isolability]
	Fault $f_i$ is structurally isolable from fault $f_j$ in a multi-mode system $\Ms$ if 
	\begin{equation}
		\eover{e_{f_i}}(\Ms \backslash \{e_{f_j}\}) \not \equiv \ffalse \ ,
	\end{equation}
	where $\Ms \backslash \{e_{f_j}\}$ is obtained by removing $e_{f_j}$ from  $\Ms$. \hfill $\blacklozenge$
\end{definition}

As an illustration, considering a single SM, one gets
\begin{equation*}
	\eover{e_{f_\text{cell}}}(\Ms \backslash \{e_{f_{i_\text{cell}}}\}) \equiv \lnot \back \land \lnot \forw \ ,
\end{equation*} 
which shows that $f_\text{cell}$ is isolable from $f_{i_\text{cell}}$  
only in the bypass modes according to Table~\ref{tab:SMswmodes}.
		
Algorithm~\ref{alg:mmiso} computes the multi-mode diagnosability $D$ of model $\Ms$ using Algorithm~\ref{alg:mmdetectability}. 
It is assumed, without loss of generality, that there is at most one fault in each equation.
		
\begin{algorithm}
	\caption{$D = \text{Diagnosability}(\Ms)$}\label{alg:mmiso}
	\begin{algorithmic}[1] %
	\State $\langle D_{(f,\texttt{NF})} \rangle_{f\in \mathcal{F}} = \text{Detectability}(\Ms)$
	\ForAll{$f \in \mathcal{F}$}
		\If{$f$ is detectable}
			\State $\langle D_{(f_i,f)} \rangle_{f_i\in \mathcal{F}} = \text{Detectability}(\Ms\setminus \{e_{f}\})$
		\Else \Comment{Avoid unnecessary \texttt{mmDM}}
			\State $\langle D_{(f_i,f)} \rangle_{f_i\in \mathcal{F}} = \langle D_{(f_i,\texttt{NF})} \rangle_{f_i\in \mathcal{F}}$  					
		\EndIf
	\EndFor
	\end{algorithmic}
\end{algorithm}
	
The diagnosability result of Algorithm~\ref{alg:mmiso} for the single SM example is shown in Table~\ref{tab:1SM_iso_matrix}. In this table, $\by$ stands for $\lnot \back \land \lnot \forw$ for clarity. 
Column NF %
indicates that all faults are detectable in all modes. The 2-by-2 block of $\ffalse$'s indicates that faults $f_\text{cell}$ and $f_{v_\text{cell}}$ are not isolable in any mode. The entries with $\by$ indicate that the faults $f_\text{cell}$ and $f_{v_\text{cell}}$ are isolable from fault $f_{i_\text{cell}}$ in the bypass modes only.
		
\begin{table}[h]
	\centering
	\caption{Multi-mode fault diagnosability for the one SM example.}
	\vspace*{-2mm}
	\label{tab:1SM_iso_matrix}
	\begin{tabular}{c|cccc}
		\hline 
		& NF  & $f_{\text{cell}}$  & $f_{v_{\text{cell}}}$ & $f_{i_{\text{cell}}}$  \\
		\hline
		$f_{\text{cell}}$   & \cellcolor{lightgray} $\ttrue$ & $\ffalse$ & $\ffalse$  & $\by$ \\
			
		$f_{v_{\text{cell}}}$ & \cellcolor{lightgray} $\ttrue$ &  $\ffalse$  &  $\ffalse$ & $\by$ \\
				
		$f_{i_{\text{cell}}}$ & \cellcolor{lightgray} $\ttrue$ &  $\by$  & $\by$ &  $\ffalse$ \\ %
		\hline
	\end{tabular}
\end{table}

\subsection{Diagnosability based on Boolean faults}	
	
In this approach, faults are modeled by the Boolean mode variables $F = (F_1, F_2, \ldots, F_n)$ such that $F_i = \ttrue$ if fault $i$ is present, $F_i = \ffalse$ otherwise. As above, $S$ is the set of Boolean variables that represent system modes; we denote by $\mathbb{M}_b$ the set of valid evaluations of the Boolean variables of $F \cup S$.
	The {\mmDM} decomposition computes, for each equation $e \in E$, a Boolean function $\eover{e} : \mathbb{M}_b \rightarrow \bool$,
where superscript + denotes once again the over-determined part of DM. The present approach similarly employs this function to assess the diagnosability of multi-mode systems. However, in contrast to the preceding methodology, the results of {\mmDM} are now functions of both system modes and faults. 
 
To define detectability, some notation is needed. Let $\Psi$ be a Boolean function of the Boolean variables in set $X$. For $x\in X$, $\Psi[x:=\ffalse / \ttrue]$ denotes the function obtained by assigning the value $\ffalse / \ttrue$ to $x$. Furthermore, for a set $X' \subseteq X$, $\Psi[X':=\ffalse / \ttrue]$ will denote the function obtained by assigning to every $x\in X'$ the value $\ffalse / \ttrue$.

\begin{definition} [Structural detectability]\label{def:Bool-detect}
	Let $\mathcal{F}$ be the set of faults in a multi-mode system $\MB$. 
	A fault $F_i$ is structurally detectable in a multi-mode model $\MB$ if 
	\begin{equation}
		\eover{e_{F_{i}}}(\MB)[\mathcal{F} := \ffalse] \not \equiv \ffalse,
	\end{equation}
	where $e_{F_i}$ is the equation related to fault $F_i$. \hfill $\blacklozenge$
\end{definition}      
Fault $F_i$ is detectable in the modes given by $\eover{e_{F_{i}}}(\MB)[\mathcal{F} := \ffalse]$, which corresponds to $\eover{e_{f_{i}}}(\Ms)$ in the fault signal approach.
 	  
Consider again the single SM example and the detectability of fault $F_{i_{\text{cell}}}$. For equation $e_{F_{i_{\text{cell}}}}$, the $\mmDM$ algorithm yields the overdetermined part 
\begin{align}\label{eq:e_plus_Fi}
	\begin{split}
	\eover{e_{F_{i_{\text{cell}}}}}(\MB) \equiv &(\lnot \back \land \lnot F_\text{cell} \land \lnot F_{i_\text{cell}} \land \lnot F_{v_\text{cell}}) \lor \\
	&(\lnot \forw \land \lnot F_\text{cell} \land \lnot F_{i_\text{cell}} \land \lnot F_{v_\text{cell}}) \lor\\
	&(\lnot \forw \land \lnot \back \land \lnot F_\text{cell} \land \lnot F_{v_\text{cell}})
	\end{split}
\end{align}
which depends on both system and fault mode variables. By assigning $\ffalse$ to all fault variables, the modes where $F_{i_\text{cell}}$ is detectable are given by
\begin{align*}
	\eover{e_{F_{i_\text{cell}}}}(\MB)[\{F_\text{cell}, F_{i_\text{cell}}, F_{v_\text{cell}}\}:= \ffalse] \equiv \\ 
	\lnot \back \lor \lnot \forw \, .
\end{align*}
This is equal to the invariant condition, therefore $F_{i_{\text{cell}}}$ is detectable in all modes.

	 \begin{definition}[Single fault structural isolability]\label{def:Bool-isolability}
		Let $\mathcal{F}$ be the set of faults in a multi-mode system $\MB$. Fault $F_i$ is structurally isolable from fault $F_j$ if 
	 	\begin{equation}
	 		\eover{e_{F_{i}}}(\MB)[ F_j:= \ttrue, \mathcal{F} \setminus \{F_j\} := \ffalse] \not \equiv \ffalse.
	 	\end{equation}
	 \end{definition}

The assignment of the fault variables establishes that model $\MB$ is valid in the mode with fault $F_j$ only. If equation $e_{F_{i}}$ is in the overdetermined part of $\MB$, the definition states that $F_i$ is isolable from $F_j$. Note that the same function $\eover{e_{F_{i}}}(\MB)$ used in Definition~\ref{def:Bool-detect} to compute detectability properties is used for computing isolability properties of the model as well. This means that $\mmDM$ only has to be run once for the complete diagnosability analysis, including both detectability and isolability analysis.    

	As an example, let us consider the single SM example and check whether $F_\text{cell}$ is isolable from $F_{i_\text{cell}}$. This is computed by the following substitution in~\eqref{eq:e_plus_Fi}:
	\begin{align*}
		&\eover{e_{F_\text{cell}}}(\MB)[F_{i_\text{cell}}:= \ttrue, \{F_\text{cell}, F_{v_\text{cell}}\}:= \ffalse] \equiv \\
		& \hspace*{4cm}\lnot \forw \land \lnot \back \ ,
	\end{align*}
	according to Defintion~\ref{def:Bool-isolability}. Thus, $F_\text{cell}$ is isolable from $F_{i_\text{cell}}$ if and only if the submodule is in a bypass mode, which is consistent with the results obtained with the fault signal methodology. 
	
From this approach, the concept of isolability can easily be extended to multiple fault isolability analysis as the following.

	\begin{definition}[Multiple fault structural isolability]
		Let $\mathcal{F}$ be the set of faults in $\MB$ and let $\Gamma \subseteq \mathcal{F}$ be a subset of faults. Fault $F_i \in \mathcal{F}$ is structurally isolable from the multiple fault $\Gamma$ if
		\begin{equation}
			\eover{e_{F_i}}(\MB)[\Gamma := \ttrue, \mathcal{F}\backslash \Gamma := \ffalse] \not \equiv \ffalse \ .
		\end{equation}
	\end{definition}

	 In the single SM model, for example, one can determine the isolability of fault $F_\text{cell}$ from the dual fault scenario including both $F_{i_\text{cell}}$ and $F_{v_\text{cell}}$ as follows:
	\begin{equation}
		\eover{e_{F_\text{cell}}}(\MB)[F_\text{cell} := \ffalse, \{F_{i_\text{cell}}, F_{v_\text{cell}}\} := \ttrue] \equiv \ffalse.
	\end{equation}
	Hence, $F_\text{cell}$ is not isolable from the double fault. 
	
	To show the applicability of these algorithms, we use them in a case study in the next section.

\section{Case study and results} \label{Results}

We assessed both modeling approaches on the battery pack model introduced in Section~\ref{sec:mmsystems}, with varying number $n$ of SMs.
The diagnosability, as given by~(\ref{diag}), is the same for both fault modeling approaches, as expected; it is summarized in Table~\ref{tab:isolability_matrix}. 
This table presents the results for $n$ SMs, as the same pattern is always observed across all SMs. This common pattern is aggregated and represented by subscript $k$ within the blue highlighted area.

The first column, highlighted in gray, shows detectability. All faults are detectable, except $f_{i_\text{pack}}$ which is undetectable if all SMs are in $\by$ mode. In cases where $f_{i_\text{pack}}$ is not detectable, other faults cannot be isolable from it, indicated in the $f_{i_\text{pack}}$ row.

It is important to note that any fault occurring within a given SM is detectable, and uniquely isolable from both pack faults and faults in any other SM. For instance, $f_{\text{cell},1}$ and $f_{\text{cell},2}$ are detectable and uniquely isolable from each other since they belong to different SMs. However, within a given SM $k$, component fault $f_{\text{cell},k}$ and sensor fault $f_{v_\text{cell},k}$ are not isolable when that particular SM is in $\by$ mode.

These findings are consistent with our earlier study \cite{paper_with_Arvind} on sensor investigation for reconfigurable battery systems. However, while that study relied on a brute force approach to systematically explore all modes and system configurations, the use of {\mmDM} has resulted in an important reduction of computational times.
\begin{table}
	\centering
	\caption{Multi-mode fault diagnosability.}
	\vspace*{-2.2mm}
	\label{tab:isolability_matrix}
	\scalebox{0.78}
	{\small
		\hfill{}
		\begin{tabular}{c|cccccc}
			\hline 
			& $NF$  & $\fltcell{k}$  & $f_{\vcell{k}}$ &  $f_{\icell{k}}$ &  $f_{\iout}$ & $f_{\vout}$\\
			\hline
			$\fltcell{k}$   & \cellcolor{lightgray}$\ttrue$ & \cellcolor{LightCyan} $\ffalse$ & \cellcolor{LightCyan} $\neg \by_k $ & \cellcolor{LightCyan}$\ttrue$ & $\ttrue$ & $\ttrue$\\
			
			$f_{\vcell{k}}$ & \cellcolor{lightgray}$\ttrue$ & \cellcolor{LightCyan} $\neg \by_k $ & \cellcolor{LightCyan}$\ffalse$ & \cellcolor{LightCyan}$\ttrue$ & $\ttrue$ & $\ttrue$\\
			
			$f_{\icell{k}}$ & \cellcolor{lightgray}$\ttrue$ & \cellcolor{LightCyan}$\ttrue$  & \cellcolor{LightCyan}$\ttrue$ & \cellcolor{LightCyan}$\ffalse$ & $\ttrue$ & $\ttrue$\\ 
			
			$f_{\iout}$     & \cellcolor{lightgray} $\neg \by_\text{all} $ & $\neg \by_\text{all} $ &  $\neg \by_\text{all} $ &  $\neg \by_\text{all} $ &  $\ffalse$ &  $\neg \by_\text{all} $ \\
			
			$f_{\vout}$ & \cellcolor{lightgray}$\ttrue$ &  $\ttrue$ &  $\ttrue$ & $\ttrue$ & $\ttrue$ & $\ffalse$\\
			\hline
	\end{tabular}}
	\hfill{}	
\end{table}
\subsection*{Fault modeling comparison}
With the Boolean fault approach, a notable advantage is the single run of {\mmDM}, followed by subsequent evaluations of various possibilities on fault modes. In contrast, within the fault signal approach, {\mmDM} must be run once per fault.

\begin{figure}[h!]
	\centering
	\includegraphics{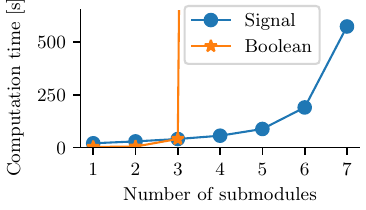}
	\vspace{-2.5mm}
	\caption{Computation time comparison for two fault modeling approaches.}
	\label{fig:computation_time_comparison}
\end{figure}

However, as demonstrated by the computational times for both approaches shown in Figure~\ref{fig:computation_time_comparison}, the Boolean fault approach leads to a rapid escalation in computational complexity, due to the higher number of Boolean variables to handle: $5n+2$ in this approach, versus $2n$ in the fault signal approach.
The computational demands for the signal fault approach are higher for a small number of SMs, but remain lower in scenarios involving more SMs, as partial evaluations of the outputs of {\mmDM} on fault modes are cheap compared to the whole multimode DM decomposition.  %

\section{Conclusion}

A novel methodology for conducting structural fault diagnostics for multi-mode systems has been introduced. An inherent challenge in diagnosing such systems arises from their ability to operate in different modes: considering all possible system configurations yields combinatorial complexity for the analysis. %

Using a multi-mode extension of the Dulmage-Mendelsohn decomposition, called {\mmDM}, an effective means is provided to address this complexity. Additionally, two distinct approaches for modeling faults are introduced, as either signals or Boolean variables. In both methodologies, the multimode extension of the definitions of detectability and isolability was introduced. To demonstrate the efficiency of our method, we applied our approaches to an illustrative example of a modular switched Li-ion battery pack with a full-bridge converter. %

Future works will aim at generalizing our approach to more complex multi-mode models, including ones where some faults can only occur in some modes. Algorithmic improvements of {\mmDM} are also in the works for improving computational times.

\bibliography{refpaper3}             %

\appendix

\end{document}